# Characterization of cough sounds using statistical analysis

Naveenkumar Vodnala[a,b,d,e], Pratap Reddy Lankireddy[a,c,d], Padma Sai Yarlagadda[a,b,d]

[a]Department of Electronics and Communication Engineering; [b]VNR Vignana Jyothi Institute of Engineering and Technology; [c]Jawaharlal Nehru Technological University; [d]Hyderabad, Telangana State, India – 500090; [e]E-mail: vnaveenkumar.vnk@gmail.com

**Abstract**

Cough is a primary symptom of most respiratory diseases, and changes in cough characteristics provide valuable information for diagnosing respiratory diseases. The characterization of cough sounds still lacks concrete evidence, which makes it difficult to accurately distinguish between different types of coughs and other sounds. The objective of this research work is to characterize cough sounds with voiced content and cough sounds without voiced content. Further, the cough sound characteristics are compared with the characteristics of speech. The proposed method to achieve this goal utilized spectral roll-off, spectral entropy, spectral flatness, spectral flux, zero crossing rate, spectral centroid, and spectral bandwidth attributes which describe the cough sounds related to the respiratory system, glottal information, and voice model. These attributes are then subjected to statistical analysis using the measures of minimum, maximum, mean, median, and standard deviation. The experimental results show that the mean and frequency distribution of spectral roll-off, spectral centroid, and spectral bandwidth are found to be higher for cough sounds than for speech signals. Spectral flatness levels in cough sounds will rise to 0.22, whereas spectral flux varies between 0.3 and 0.6. The Zero Crossing Rate (ZCR) of most frames of cough sounds is between 0.05 and 0.4. These attributes contribute significant information while characterizing cough sounds.

# 1. Introduction

World Health Organization (WHO) states that Chronic Respiratory Diseases (CRDs) affect the airways and other structures of the lungs, with some common examples being Chronic Obstructive Pulmonary Disease (COPD), asthma, occupational lung diseases, and pulmonary hypertension. These are the leading causes of death in developing and lower-income countries. CRDs are not curable, but early detection and regular monitoring can help manage symptoms and improve daily life for those living with these conditions. Spirometry is the most common lung function test used to measure airflow limitation. However, it is challenging to access spirometry equipment in remote areas, and some people may experience adverse effects from the test. Chest X-rays and CT scans help in diagnosis by providing images of the lungs, but excessive exposure to X-rays is harmful. A replacement or supportive mechanism is needed to overcome the challenges associated with spirometry and Radiology [1].

Cough is a primary symptom of these respiratory diseases. The sound of a cough specifies the pathophysiological mechanisms of coughing, providing information on the structural nature of the respiratory airways. The basic cough sound pattern includes an initial burst of sound, a noisy interval, and a second burst of sound. Pathological events such as airway narrowing, bronchoconstriction, fibrosis, and inflammation alter the cough sound pattern [2]. Few healthcare professionals categorized cough sounds as cough alone, cough with mucus, cough with wheeze, cough with wheeze and mucus, and also found that coughs with mucus have longer second phases [3]. Cough is initiated by sensory receptors in the larynx and lower respiratory tract that sends signals to the brainstem. The sensitivity and pattern of coughing are influenced by C-fibre receptors and Rapidly Adapting Receptors (RARs) [4]–[6].

The respiratory diagnosis system presented in [7] utilizes feature vectors from the time domain, frequency domain, and mixed domain to achieve high accuracy. The coughing sound is analyzed using spectral analysis to identify changes in frequency bands [8]. FFT spectral analysis [3] is used to distinguish acoustic differences between coughs with and without mucus in voluntary cough sounds. Cough sounds are quantified using cough epochs and measured cough intensity [9] using cough sound power, peak energy, and mean energy. The study in [10] has identified two frequency bands in cough spectra based on their low-tone prominence, harmonicity, and high-frequency components. Obstructive diseases and restrictive diseases [1] are also differentiated using cough sounds. A comparison of acoustic features of cough sounds between pneumonia and non-pneumonia groups is done [11] by analyzing time, frequency, psychoacoustics, and energy. An investigation [12] separated cough sounds into intrinsic mode-function components using the empirical-mode decomposition method based on frequency bands. As part of an investigation in [13], the authors distinguished cough sounds from non-cough sounds but did not characterize cough sounds. However, similar cough sounds produced by different respiratory illnesses, and co-morbidities [14] in many patients, which must be considered in real-world situations. Audio features [15] are found to be helpful to build machine-learning classifiers for cough-based diagnosis engines. Artificial Intelligent (AI) based approaches and deep learning models [16]–[18] are proposed for cough sound detection and differentiation of respiratory diseases. Most of the machine-learning techniques utilize automated extraction of time-frequency cough features [19], [20] for the diagnosis of COVID-19.

However, concrete evidence is necessary to substantiate the characterization of cough sounds. Misinterpretation of cough sounds will have significant consequences, including erroneous outcomes and irretrievable miscalculations in future experiments. It is important to identify and differentiate between different types of coughs and other acoustics. To address these research gaps, an analysis of cough sounds is conducted to characterize them using spectral and time domain attributes.

## 2. Material and methods

Cough sounds and audible voiced signals are differentiated by the way of vibration of vocal folds [21]. The spectral and time domain characterization of these sounds will be employed by extracting features related to the respiratory system, glottal information, and voice model [22]. Spectral roll-off, spectral entropy, spectral flatness, spectral flux, zero crossing rate, spectral centroid, and spectral bandwidth attributes [23], [24] are considered for cough sound characterization in this research work. The underlying respiratory conditions will play a major role to distinguish various types of cough [25], [26]. In the present work, speech sounds are considered for reference information while characterizing sounds.

A coughing sound with content such as a grunt or a groan embedded in it is found with specific characteristics that include voiced components. In such a scenario, the cough sound is titled "coughing sound with voiced content" which is presented in Figure 1(a). A coughing sound which is produced primarily by the expulsion of air from the lungs does not possess audible vocal components, in which case, those are titled "cough sound without voiced content" and presented in Figure 1(b).

In the present experiment, recordings used for characterization consist of 20 recordings of coughing sounds with voiced content and coughing sounds without voiced content each type. The duration of these recordings ranges from 110 msec to 420 msec, and they are captured in noise-free environments. These cough sounds are a mixture of samples collected from both healthy and unhealthy, male and female, adults and youngsters. The statistical analysis is done using, minimum (min), maximum (max), mean, 25% of the median (med_25), median, 75% of the median (med_75), and standard deviation (std). The sampling rate for each of the sounds is 22050 Hz. Each record's time series is divided into segments, converted into frames of 512 samples (23 msec), and each frame is examined with a 50% frame overlap.

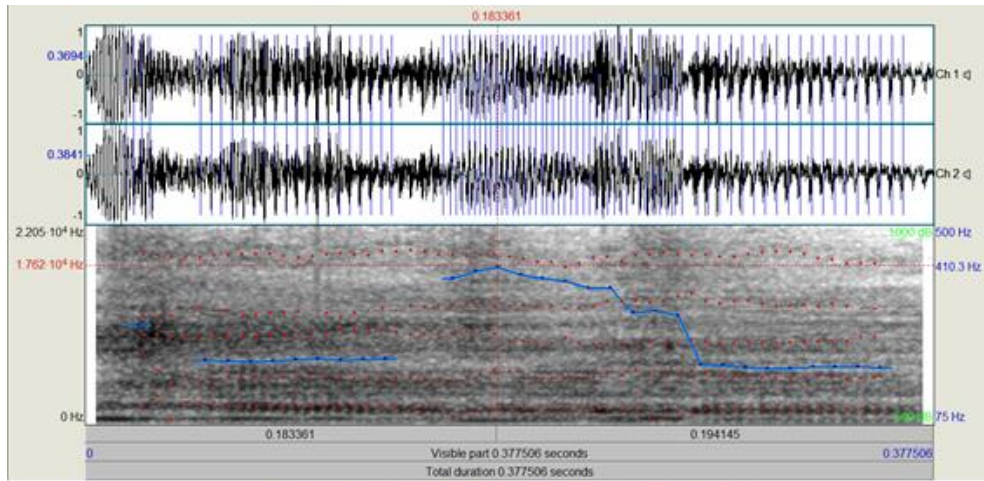

(a) A coughing sound with voiced content

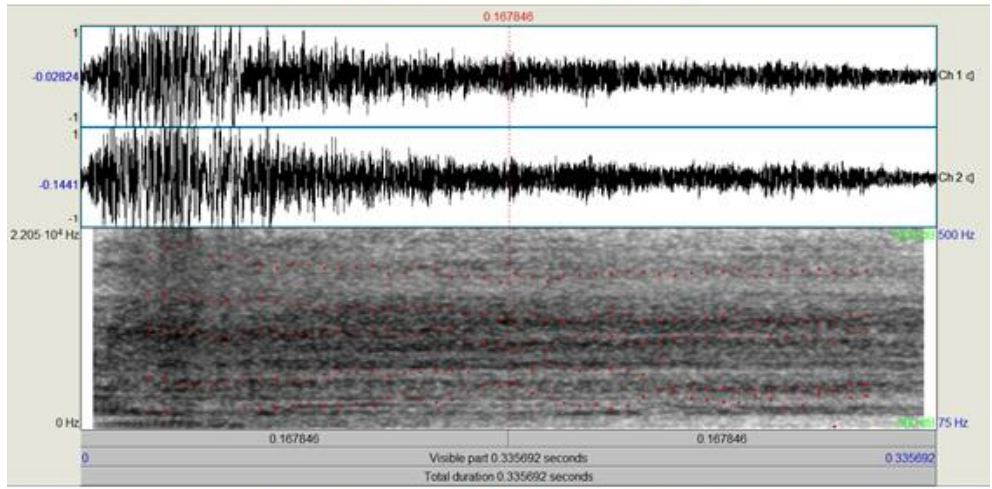

(b) A coughing sound without voiced content

**Figure 1. A coughing sound (a) with voiced content (b) without voiced content.**

## 2.1 Spectral roll-off

The roll-off frequency represents the center frequency of a specific bin in the spectrogram such that a minimum roll percent of the energy in the spectrum of the frame is contained within this bin and all the bins below it. In this study, 85% is taken as the roll percent spectral roll-off is measured using equation (1).

$$\text{SRO} = f(i) \text{ for which } \frac{\sum_{j=1}^{i} p(j)}{\sum_{j=1}^{N} p(j)} \geq roll\_percent \tag{1}$$

where:

SRO is spectral roll-off frequency

$f(i)$ is the center frequency of the $i^{th}$ frame

$p(j)$ is the magnitude-squared of the frequency response of the $j^{th}$ bin

N is the number of frequency bins

roll_percent = 85%

The spectral roll-off of cough sounds is shifted toward higher frequencies due to the explosive nature of the sound. Coughing sounds with voiced content have spectral roll-off varying from 1378 Hz to 9216 Hz, with a mean of 4451 Hz, which is presented in Figure 2(a). The coughing sounds without voiced content have spectral roll-off varying from 1808 Hz to 9819 Hz, with a mean of 5705 Hz, which is presented in Figure 2(b).

## 2.2. Spectral entropy

Spectral entropy is a measure of the randomness or uncertainty of the energy distribution in a signal's frequency domain. Higher entropy indicates higher uncertainty and a more chaotic system. Spectral entropy is calculated using equation (2).

$$H(X) = - \sum_{i=1}^{i=N} P(f_i) log2(P(f_i)) \tag{2}$$

where:

H(X) is the spectral entropy of the signal X

N is the number of frequency bins in the signal's spectrum

$P(f_i)$ is the normalized energy of the ith frequency bin

Cough sounds tend to have higher spectral entropy, indicating more complexity and less tonality. The entropy values are higher in the initial burst of cough sounds, indicating high uncertainty, which is presented in Figures 2(c) and 2(d).

*2.3. Spectral flatness*

Spectral flatness provides information about how evenly the energy of a sound is distributed across different frequencies. The spectral flatness is obtained as the ratio of the geometric mean of the power spectrum to its arithmetic mean using equation (3).

$$\text{SFT} = \frac{\sqrt[N]{\prod_{n=0}^{N-1} p(n)}}{\frac{1}{N}\sum_{n=0}^{N-1} p(n)} \qquad (3)$$

where:

SFT is spectral flatness

$p(n)$ is the magnitude-squared of the frequency response of the $n^{th}$ bin

N is the number of frequency bins

Figure 2(e) demonstrates that coughing sounds with voiced content have spectral flatness, predominantly near '0', with a maximum value of 0.047. This indicates that harmonic components dominate and there is less noise since these sounds are generated by periodic vocal fold vibrations. The coughing sounds without voiced contents have spectral flatness farther away from '0', with a maximum value of 0.22, indicating more noise is presented in Figure 2(f). This is because unvoiced cough sounds lack periodic vocal fold vibrations.

*2.4. Spectral flux*

Spectral flux is measured as the squared difference between the normalized magnitudes of the spectra of the two successive frames and measures the spectral change between two successive frames using equation (4). Spectral flux points to sudden change in spectral magnitudes.

$$SFL_{(i,i-1)} = \sum_{k=1}^{k=WL}(E_i(k) - E_{i-1}(k))^2 \qquad (4)$$

Where:

SFL is the spectral flux of $i^{th}$ and $(i-1)^{th}$ frames

$E_i(k)$ is the normalized magnitude of the $k^{th}$ bin in an $i^{th}$ frame.

*WL* is the length of the frame.

The power spectrum of cough sounds exhibits a rapid change at the onset of coughing, making spectral flux a useful tool for detecting the onset of coughing. Figures 2(g) and 2(h) provide the spectral flux for both coughing sounds with voiced content and coughing sound without voiced content, having a mean value of 0.43. These findings indicate that the spectral content of cough sounds changes rapidly over time, which is given by the dynamic nature of cough sounds, varying significantly in terms of frequency content and intensity.

*2.5. Zero Crossing Rate (ZCR)*

The ZCR is a measure of the number of times the sign of signal changes from positive to negative or vice versa in a given time interval, which is given in equation (5).

$$\text{ZCR} = \frac{1}{N}\sum_{n=1}^{n=N-1} |s[n]| > \varepsilon \tag{5}$$

Where:

ZCR is the Zero Crossing Rate

N is the length of the signal

s[n] is the value of the signal at time step n

ε is a small threshold, and the summation is taken over all time steps n from 1 to N-1

ZCR for the coughing sound without voiced content is higher, with a maximum value of 0.51 and with a mean of 0.22 compared to the coughing sound with voiced content, which is demonstrated in Figures 2(i) and 2(j). The ZCR for coughing sounds with voiced content is lower comparatively, with a mean of 0.1 and a maximum value of 0.23.

*2.6. Spectral centroid*

The spectral centroid measures the center of mass of a sound's frequency spectrum and provides information about the "brightness" of the sound and its location in the frequency domain using equation (6). It gives the frequency band where most of the energy is concentrated.

$$C_i = \frac{\sum_{k=1}^{k=WL}(f(k)P(k))}{\sum_{k=1}^{k=WL} P(k)} \tag{6}$$

Where:

$C_i$ is the centroid of $i^{th}$ frame

f(k) is the frequency of the $k^{th}$ bin in the frequency spectrum

P(k) is the power of the $k^{th}$ bin

*WL* is the length of the frame

Cough sounds have a higher spectral centroid because cough sounds typically have more energy at higher frequencies, due to the burst of air produced during a cough, which is presented in Figures 2(k) and 2(l). The spectral centroid of a cough with voiced contents has a range from 1045 Hz to 3996 Hz, with a mean of 2154 Hz, whereas the spectral centroid of a cough without voiced contents has a range from 907 Hz to 5397 Hz, with a mean of 3124 Hz.

*2.7. Spectral Bandwidth*

Spectral bandwidth is used to describe the width of a frequency band or the distribution of energy within a frequency band. It is measured as the spread of the power spectrum around its centroid using equation (7). The bandwidth will change over time within a single cough sound, which is indicative of changes in the properties of the underlying airway or lung tissue. Figures 2(m) and 2(n) depict the spectral bandwidth with a voiced cough sound and an unvoiced cough sound. The spectral bandwidth of a coughing sound with voiced contents has a range from 1273 Hz to 3605 Hz, with a mean of 2286 Hz, whereas the spectral centroid of a coughing sound without voiced contents has a range from 1307 Hz to 3760 Hz, with a mean of 2348 Hz.

$$SBW_i = \frac{\sum_{k=1}^{k=WL}(|f(k)-C_i|P(k))}{\sum_{k=1}^{k=WL} P(k)} \tag{7}$$

Where:

$SBW_i$ is the spectral bandwidth of $i^{th}$ frame

$C_i$ is the spectral centroid of $i^{th}$ frame

f(k) is the frequency of the $k^{th}$ bin in the frequency spectrum

P(k) is the power of the $k^{th}$ bin, and *WL* is the length of the frame

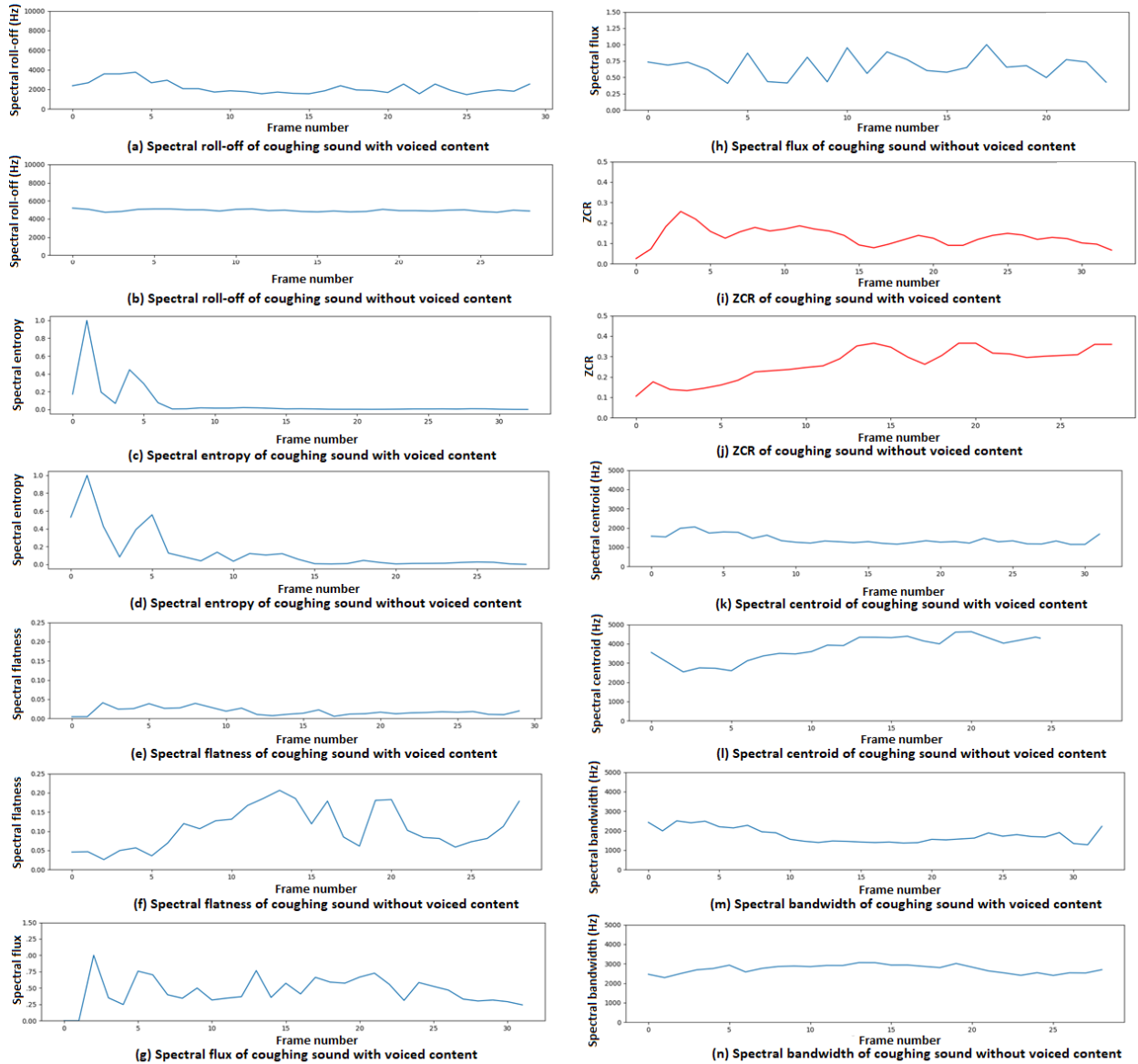

**Figure 2.** Spectral roll-off of (a) coughing sound with voiced content (b) coughing sound without voiced content; Spectral entropy of (c) coughing sound with voiced content (d) coughing sound without voiced content; Spectral flatness of (e) coughing sound with voiced content (f) coughing sound without voiced content; Spectral flux of (g) coughing sound with voiced content (h) coughing sound without voiced content; Zero Crossing Rate of (i) coughing sound with voiced content (j) coughing sound without voiced content; Spectral centroid of (k) coughing sound with voiced content (l) coughing sound without voiced content; Spectral bandwidth of (m) coughing sound with voiced content (n) coughing sound without voiced content.

## 3. Results and Discussion

Tables 1, 2, and 3 provide statistical analyses of cough with voiced content, cough without voiced content, and speech signals, respectively. In the experimental setup, 20 speech recordings are considered along with cough sound recordings. These speech signals are random utterances of one or two words, such as "He Like," "Buying," "Don't Ask," "Money," etc. are randomly chosen from the Texas Instruments Massachusetts Institute of Technology (TIMIT) speech database.

**Table 1. Statistical analysis of cough with voiced content**

| Attributes | min | max | Mean | med_25 | median | med_75 | Std |
|---|---|---|---|---|---|---|---|
| Spectral Roll-off (in Hz) | 1378 | 9216 | 4451 | 2670 | 4392 | 5943 | 1949 |
| Spectral Entropy | 0 | 1 | 0.067 | 0.001 | 0.005 | 0.021 | 0.174 |
| Spectral Flatness | 0 | 0.047 | 0.012 | 0.003 | 0.01 | 0.018 | 0.011 |
| Spectral Flux | 0 | 1 | 0.409 | 0.293 | 0.402 | 0.551 | 0.225 |
| Zero Crossing Rate | 0.025 | 0.225 | 0.1 | 0.061 | 0.086 | 0.131 | 0.046 |
| Spectral Centroid (Hz) | 1045 | 3996 | 2154 | 1542 | 2122 | 2619 | 700 |
| Spectral Bandwidth (Hz) | 1273 | 3605 | 2286 | 1864 | 2276 | 2680 | 527 |

**Table 2. Statistical analysis of cough with unvoiced content**

| Attributes | min | max | Mean | med_25 | median | med_75 | Std |
|---|---|---|---|---|---|---|---|
| Spectral Roll-off (in Hz) | 1808 | 9819 | 5705 | 4392 | 5555 | 7321 | 1961 |
| Spectral Entropy | 0 | 1 | 0.079 | 0.003 | 0.023 | 0.076 | 0.15 |
| Spectral Flatness | 0 | 0.221 | 0.057 | 0.009 | 0.036 | 0.094 | 0.057 |
| Spectral Flux | 0 | 1 | 0.438 | 0.324 | 0.45 | 0.552 | 0.222 |
| Zero Crossing Rate | 0.035 | 0.51 | 0.222 | 0.144 | 0.230 | 0.305 | 0.097 |
| Spectral Centroid (Hz) | 907 | 5397 | 3124 | 2473 | 3229 | 3730 | 917 |
| Spectral Bandwidth (Hz) | 1307 | 3760 | 2348 | 1807 | 2470 | 2792 | 574 |

**Table 3. Statistical analysis of speech signals**

| Attributes | min | max | Mean | med_25 | median | med_75 | Std |
|---|---|---|---|---|---|---|---|
| Spectral Roll-off (in Hz) | 861 | 3445 | 1770 | 1335 | 1636 | 2153 | 597 |
| Spectral Entropy | 0 | 1 | 0.062 | 0 | 0 | 0.01 | 0.18 |
| Spectral Flatness | 0 | 0 | 0 | 0 | 0 | 0 | 0 |
| Spectral Flux | 0.0 | 1 | 0.304 | 0.148 | 0.258 | 0.44 | 0.221 |
| Zero Crossing Rate | 0.02 | 0.102 | 0.060 | 0.051 | 0.06 | 0.07 | 0.016 |
| Spectral Centroid (Hz) | 583 | 1626 | 1019 | 867 | 994 | 1152 | 221 |
| Spectral Bandwidth (Hz) | 381 | 1349 | 745 | 565 | 719 | 874 | 218 |

From the experimental results, it is observed that the mean spectral roll-off for cough sound frames is 4451 Hz and 5705 Hz, whereas, for speech signals, it is 1770 Hz. Furthermore, the spectral roll-off of cough sounds extended to higher frequencies, up to 9819 Hz, whereas speech signals are limited to 3445 Hz. The histogram presented in Figure 3(a) indicates that most of the frames for coughing sounds with voiced content are distributed between 1550 Hz and 1950 Hz and between 4100 Hz and 4700 Hz. However, because these cough sounds contained voiced content, some spectral roll-off values also appeared at lower frequencies. For coughing sounds without voiced content, as presented in Figure 3(b) the spectral roll-offs are primarily distributed between 4200 Hz and 5200 Hz, and between 7000 Hz and 7800 Hz, with all roll-off values appearing at high frequencies. In speech signals, which is demonstrated in Figure 3(c) the majority of the frames are distributed between spectral roll-off values of 1100 Hz and 1500 Hz, which are lower frequencies. Increased spectral roll-off during the compression phase, indicating strong expulsion and distinguishing coughs from other acoustics like speech.

The statistical analysis reveals the fact that voiced cough sounds exhibit spectral flatness levels ranging from 0 to 0.047, whereas cough sounds without voiced content possess spectral flatness values ranging from 0 to 0.22. This suggests that the energy in unvoiced cough sounds is moderately distributed across the spectrum's frequency ranges, as these sounds are aperiodic within a given frame. Speech, on the other hand, has nearly zero spectral flatness values but not zero across all frames when compared to cough sounds because speech has an almost periodic nature within the given frames. The spread of spectral flatness values is presented in Figures 4(a), 4(b), and 4(c) using a histogram. The higher spectral flatness in cough sounds compared to speech indicates the presence of airway obstruction or turbulence caused by inflammation or mucus.

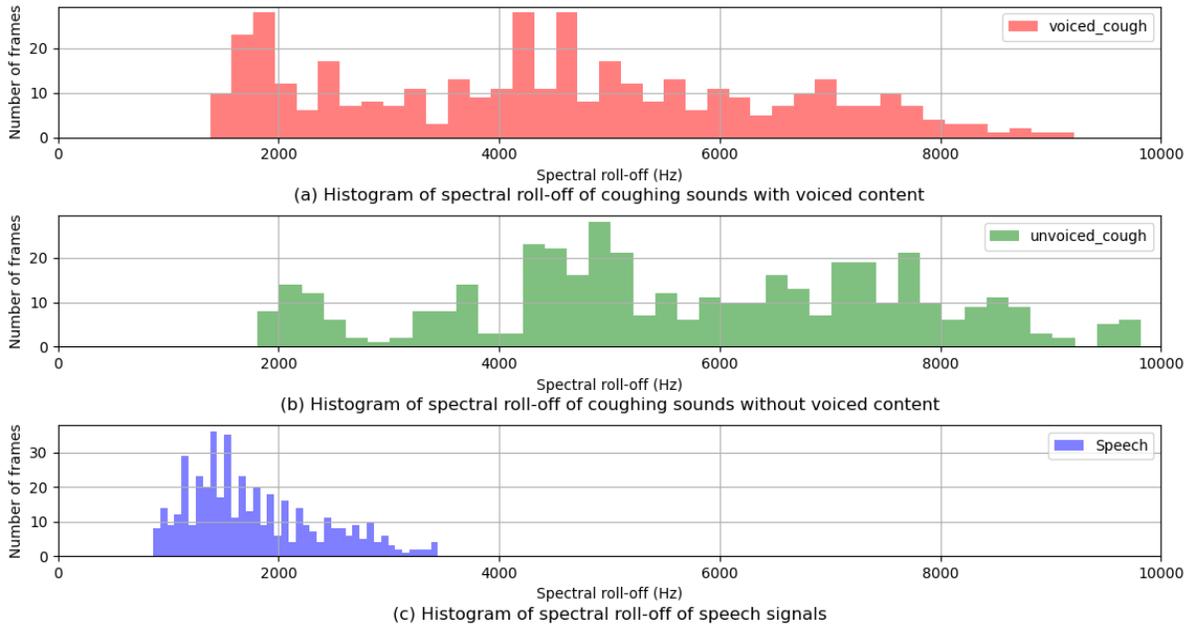

**Figure 3. Histogram of spectral roll-off of (a) coughing sounds with voiced content (b) coughing sounds without voiced content (c) speech signals.**

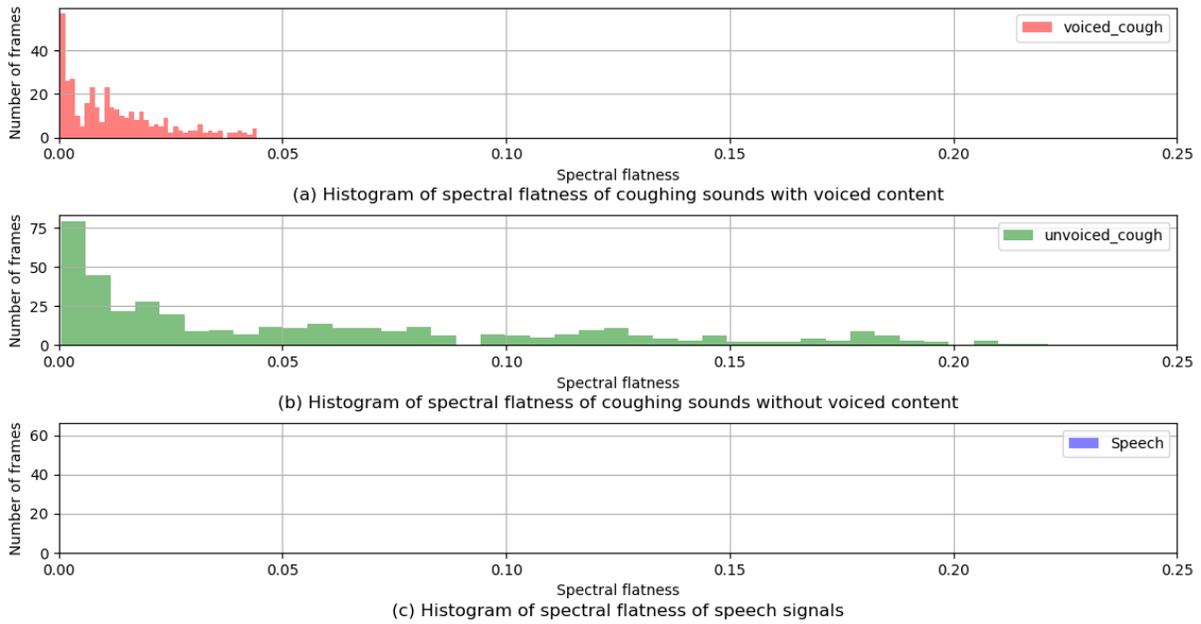

**Figure 4. Histogram of spectral flatness of (a) coughing sounds with voiced content (b) coughing sounds without voiced content (c) speech signals.**

The histogram representation in Figures 5(a), 5(b), and 5(c) shows the distribution of spectral flux values for cough sounds and speech signals. The majority of frames in cough sounds, whether they are with voiced content or without voiced content, have spectral flux values between 0.3 and 0.6. On the other hand, the majority of frames in speech signals have spectral flux values between 0.05 and 0.3, indicating that speech signals have relatively lower spectral variation compared to cough sounds. A high spectral flux in cough sounds compared to speech indicate rapid changes in airway obstruction caused by various infections or physiological transformations. This causes airway narrowing and increased turbulence during cough production.

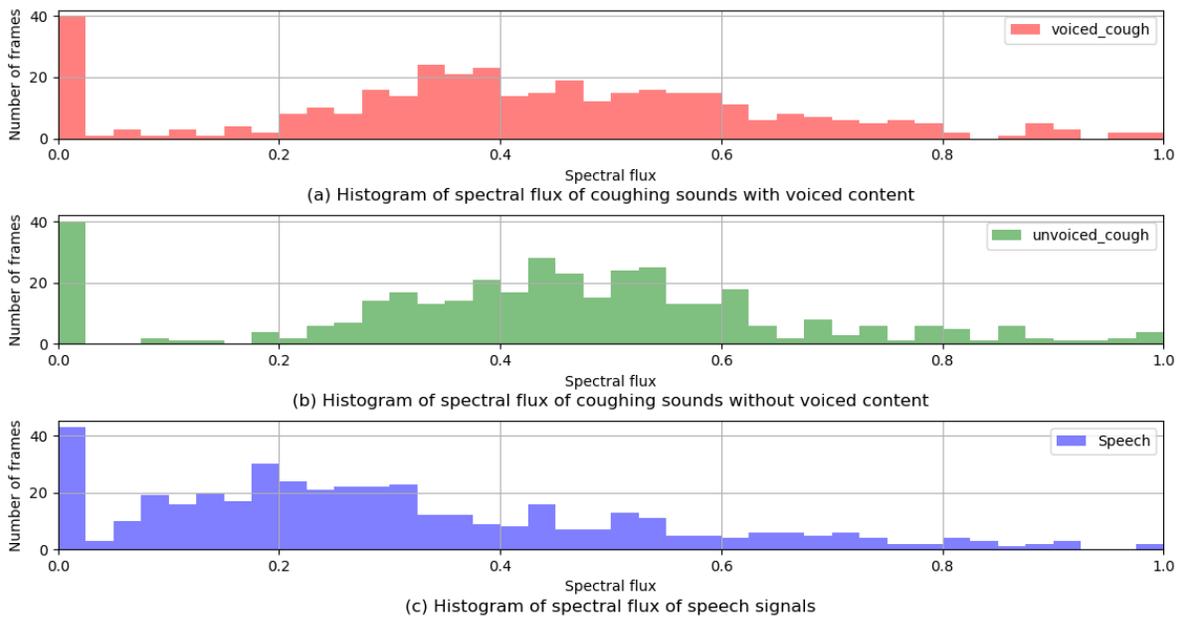

**Figure 5. Histogram of the spectral flux of (a) coughing sounds with voiced content (b) coughing sounds without voiced content (c) speech signals.**

From the results of statistical analysis, the normalized maximum zero crossing rate values for voiced cough sounds, unvoiced cough sounds, and speech are identified as 0.23, 0.51, and 0.1, with mean values of 0.1, 0.22, and 0.06, respectively, which is illustrated in Figures 6(a), 6(b), and 6(c). The histogram reveals that most frames of unvoiced cough sounds are distributed over zero crossing rate values of 0.05 and 0.4, indicating that unvoiced cough sounds have a higher zero crossing rate compared to voiced cough sounds and speech signals. The spread of zero crossing rate values for voiced cough and speech signals is between 0.05 and 0.2.

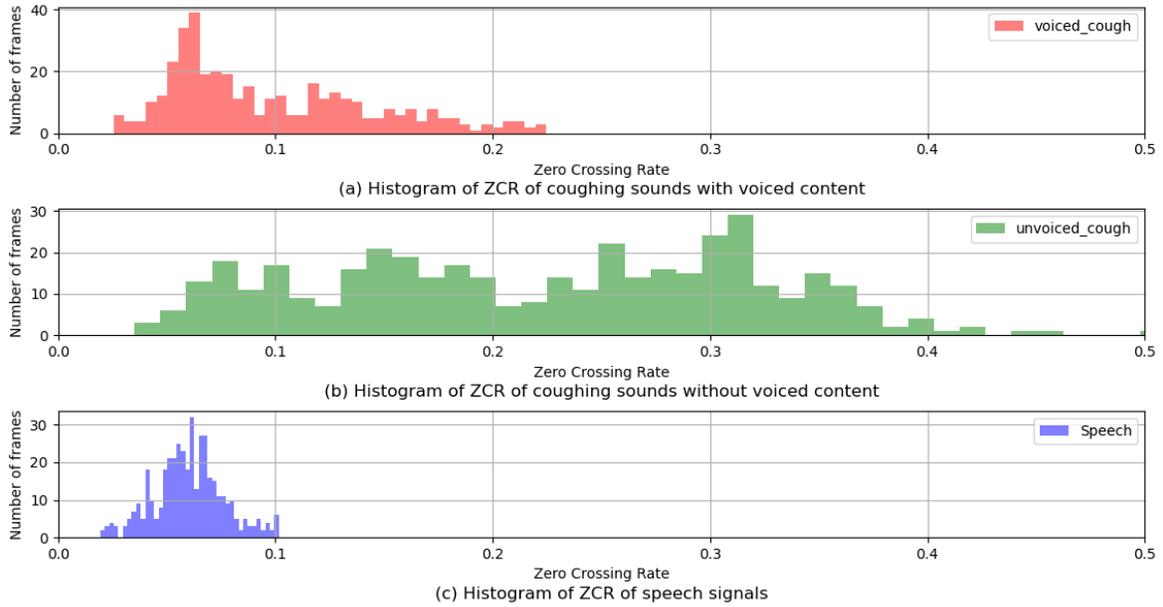

**Figure 6. Histogram of Zero Crossing Rate of (a) coughing sounds with voiced content (b) coughing sounds without voiced content (c) speech signals.**

The spread of spectral centroid values for speech signals between 583 Hz and 1626 Hz, which is lower than that of voiced and unvoiced cough sounds, which are between 1045 Hz and 3996 Hz and 907 Hz and 5397 Hz, respectively. The mean value of spectral centroid for voiced and unvoiced cough frames are 2154 Hz and 3124 Hz, while for speech signals, it is 1019 Hz. During the expulsion phase, the glottis opens, and a burst of air is expelled from the lungs, resulting in an increase in the amplitude and duration of the cough sound. And also the spectral centroid of the cough sound varies depending on the location of the cough in the respiratory tract. The histograms in Figures 7(a), 7(b), and 7(c) present the distribution of the majority of frames for each sound type.

From the statistical results, the spectral bandwidth values of speech signals range from 381 Hz to 1349 Hz, which is narrow compared to the range of voiced and unvoiced cough sounds that range from 1307 Hz to 3760 Hz and 1273 Hz to 3605 Hz, respectively. The mean values for spectral bandwidth of voiced and unvoiced cough frames are 2286 Hz and 2348 Hz, respectively, while the mean value for speech signals is 745 Hz. Based on the histogram in Figures 8(a) and 8(b), most frames in cough sounds have a wider frequency spread, ranging from 1300Hz to 3200Hz, compared to speech signals. The majority of frames in speech signals, presented in Figure 8(c) have a narrow frequency spread, ranging from 400Hz to 1000Hz. The narrow spread in speech signals is due to the vibrations of the vocal cords and resonances of the vocal tract, which tend to produce energy in a narrower frequency range.

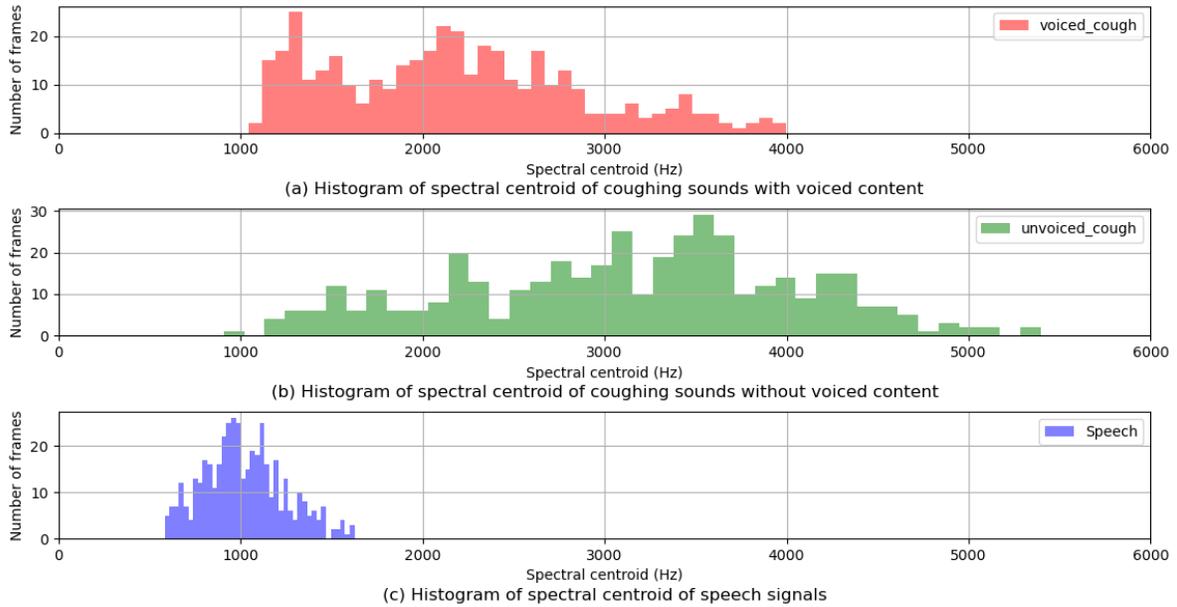

**Figure 7. Histogram of the spectral centroid of (a) coughing sounds with voiced content (b) coughing sounds without voiced content (c) speech signals.**

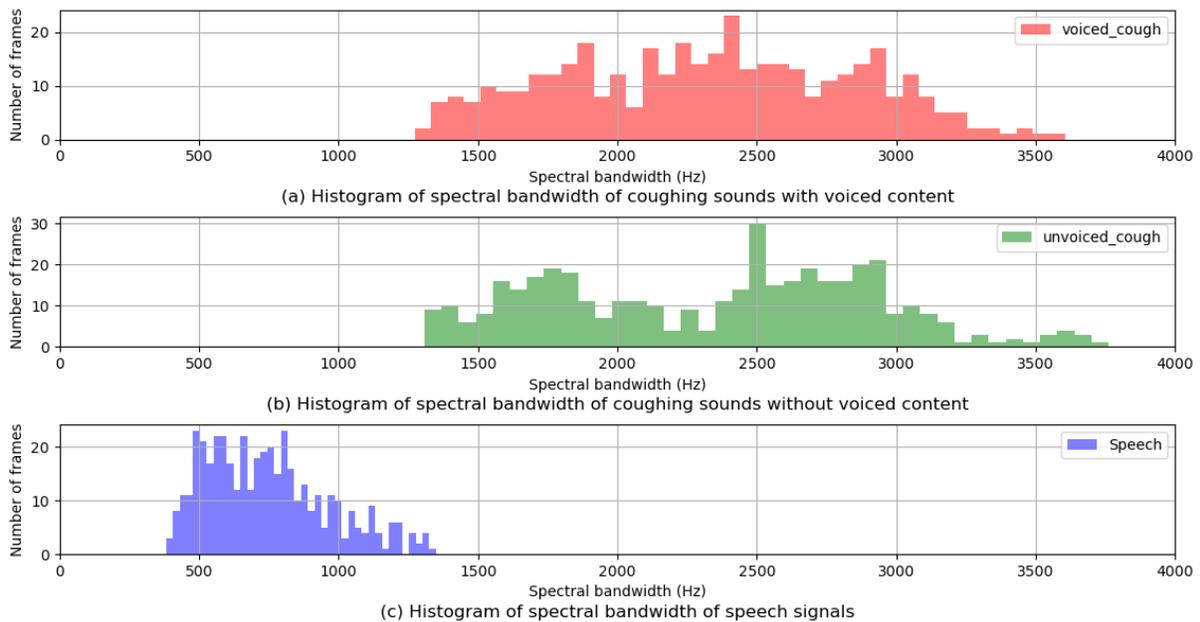

**Figure 8. Histogram of the spectral bandwidth of (a) coughing sounds with voiced content (b) coughing sounds without voiced content (c) speech signals.**

Representing the attributes of cough sounds as a range instead of a single value provides a precise and comprehensive representation of the cough sound characteristics. The spectral roll-off, spectral centroid, and spectral bandwidth attributes are employed to distinguish cough sounds from speech and simplify the description of cough sounds. The attributes of spectral flatness, spectral flux, and ZCR are useful in the clear classification of speech and cough sounds. However, it is observed that spectral entropy does not exhibit significant variation among considered sounds. These statistics are useful for the identification and treatment of respiratory diseases and for the development of diagnostic systems.

## 4. Conclusion

In this paper, the characterization of cough sounds using spectral and time domain attributes is presented. Spectral roll-off, spectral flatness, spectral flux, zero crossing rate, spectral centroid, and spectral bandwidth are important attributes utilized to describe cough sounds. Statistical analysis is performed on cough sounds with voiced content and those without, and speech sounds are used as a reference for characterizing the sounds. The analysis involved calculating the minimum, maximum, mean, median, 25% of the median, 75% of the median, and standard deviation. The results demonstrate that the considered attributes of cough sound exhibit distribution patterns in a distinctive manner. Additionally, a histogram analysis is conducted to compare cough sounds with speech signals, revealing that the considered attributes of cough sounds are distributed in closer proximity to each other compared to speech signals. To achieve efficient characterization of cough sounds, it is important to consider a group of attributes rather than a single attribute. While experiments in this paper considered significant attributes, extending the research to include additional attributes such as Mel Frequency Cepstral Coefficients (MFCCs), Linear Prediction Cepstrum Coefficient (LPCCs), etc., may further improve the characterization of cough sounds.

**Funding**

This research did not receive any specific grant from funding agencies in the public, commercial, or not-for-profit sectors.

**Conflicts of interest**

The authors declare no conflicts of interest.

**Acknowledgment**

I would like to thank Mr. S.V.N. Narayana Rao, the Director of Salcit Technologies Pvt. Ltd. in Hyderabad, Telangana, India, for his provision of authenticated cough sounds of different respiratory conditions, which are labeled by expert doctors. Moreover, I am grateful for his valuable insights and suggestions as a scientific advisor to the research team.